\documentclass[
10pt, % Main document font size
a4paper, % Paper type, use 'letterpaper' for US Letter paper
oneside, % One page layout (no page indentation)
headinclude,footinclude, % Extra spacing for the header and footer
BCOR5mm, % Binding correction
]{scrartcl}

\usepackage[
nochapters, % Turn off chapters since this is an article        
beramono, % Use the Bera Mono font for monospaced text (\texttt)
eulermath,% Use the Euler font for mathematics
pdfspacing, % Makes use of pdftex’ letter spacing capabilities via the microtype package
dottedtoc % Dotted lines leading to the page numbers in the table of contents
]{classicthesis} % The layout is based on the Classic Thesis style

\usepackage{arsclassica} % Modifies the Classic Thesis package

\usepackage[T1,T2A]{fontenc} % Use 8-bit encoding that has 256 glyphs

\usepackage[utf8]{inputenc} % Required for including letters with accents

\usepackage{graphicx} % Required for including images
\graphicspath{{Figures/}} % Set the default folder for images

\usepackage{enumitem} % Required for manipulating the whitespace between and within lists

\usepackage{lipsum} % Used for inserting dummy 'Lorem ipsum' text into the template

\usepackage{subfig} % Required for creating figures with multiple parts (subfigures)

\usepackage{amsmath,amssymb,amsthm} % For including math equations, theorems, symbols, etc

\usepackage{varioref} % More descriptive referencing

\usepackage[numbers]{natbib} % So that \citep is recognized

\theoremstyle{definition} % Define theorem styles here based on the definition style (used for definitions and examples)

\theoremstyle{plain} % Define theorem styles here based on the plain style (used for theorems, lemmas, propositions)

\theoremstyle{remark} % Define theorem styles here based on the remark style (used for remarks and notes)

\hypersetup{
colorlinks=true, breaklinks=true, bookmarks=true,bookmarksnumbered,
urlcolor=webbrown, linkcolor=RoyalBlue, citecolor=webgreen, % Link colors
pdftitle={}, % PDF title
pdfauthor={\textcopyright}, % PDF Author
pdfsubject={}, % PDF Subject
pdfkeywords={}, % PDF Keywords
pdfcreator={pdfLaTeX}, % PDF Creator
pdfproducer={LaTeX with hyperref and ClassicThesis} % PDF producer
} % Include the structure.tex file which specified the document structure and layout

\deffootnote{1.5em}{1em}{\thefootnotemark\hspace{0.1em}}
\renewcommand*{\thefootnotemark}{\normalfont\arabic{footnote}}  % ← NEU: Fix für fehlende Nummer

\makeatletter
\renewcommand*{\@makefnmark}{%
	\raisebox{3pt}[0pt][0pt]{\@textsuperscript{%
			\normalfont\hyper@linkstart{link}{footnote.\thefootnote}\arabic{footnote}\hyper@linkend%
	}}%
}
\makeatother

\usepackage{hyperref}

\title{\normalfont\spacedallcaps{On the utility of textbooks in understanding the Renaissance of General Relativity}} % The article title

\author{\spacedlowsmallcaps{Bernadette Lessel*}} % The article author(s) - author affiliations need to be specified in the AUTHOR AFFILIATIONS block

\date{\today} % An optional date to appear under the author(s)

\begin{document}

%----------------------------------------------------------------------------------------
%	HEADERS
%----------------------------------------------------------------------------------------

\renewcommand{\sectionmark}[1]{\markright{\spacedlowsmallcaps{#1}}} % The header for all pages (oneside) or for even pages (twoside)
\lehead{\mbox{\llap{\small\thepage\kern1em\color{halfgray} \vline}\color{halfgray}\hspace{0.5em}\rightmark\hfil}} % The header style

\pagestyle{scrheadings} % Enable the headers specified in this block

%----------------------------------------------------------------------------------------
%	TABLE OF CONTENTS & LISTS OF FIGURES AND TABLES
%----------------------------------------------------------------------------------------

\maketitle % Print the title/author/date block

\setcounter{tocdepth}{2} % Set the depth of the table of contents to show sections and subsections only

\tableofcontents % Print the table of contents

%\listoffigures % Print the list of figures

%\listoftables % Print the list of tables

%----------------------------------------------------------------------------------------
%	ABSTRACT
%----------------------------------------------------------------------------------------

\section*{Abstract}
This paper argues for the historical utility of general relativity textbooks written during, or shortly before, the Renaissance of general relativity. More so than research papers, textbooks offer a narrative account of the field—one that may differ substantially from the accounts of other authors addressing the same subject. Textbooks on general relativity therefore constitute a distinctive form of testimony to the intellectual forces that shaped general relativity, particularly during its reinvention period from 1955 to the mid-1970s.
A second pillar of this argument is the observation that nearly all principal investigators at the research centers previously identified by Lalli et al.—centers that functioned as intellectual epicenters of the Renaissance of general relativity—eventually produced a textbook associated with their own school of thought. To situate this observation within the broader development of the field, the paper presents a concise history of general relativity textbooks from the theory’s inception in 1915–16 to the 1980s. This history illustrates one way in which textbooks can serve as sources for reconstructing the intellectual dynamics of the Renaissance of general relativity.

%----------------------------------------------------------------------------------------
%	AUTHOR AFFILIATIONS

\let\thefootnote\relax\footnotetext{* \textit{Lichtenberg Group for History and Philosophy of Physics, University of Bonn}}

%----------------------------------------------------------------------------------------

\newpage % Start the article content on the second page, remove this if you have a longer abstract that goes onto the second page

%----------------------------------------------------------------------------------------
%	INTRODUCTION
%----------------------------------------------------------------------------------------

\section{Introduction}
Scientific textbooks comprise an interesting scholarly genre. For they, much more than stand alone research papers, narrate a story. The format --including its size, its pedagogical ambition, its general self-understanding and the necessity to begin with preliminaries-- provides space and a framework for the author(s) to narrate their own view of the topic at hand and their perspective on the corresponding field, literally from a beginning to an end. Elements of what will henceforth be called "the story" of the textbook are not just the words written in natural language. The story is comprised of the entire content of the book, including all the technicalities. It is comprised of the way (and which!) notions and concepts are introduced, of the central epistemological concerns, of how it is being argued for or against this or that point of view, validity of an argument, or the useful- and correctness of a theorem, or an assumption. In sum, the story of a textbook is comprised of the entirety of how the topic is being developed and of the thematic emphasis that is being placed. And of course, each story has its own style in which it is being narrated.

Applied to the special case of general relativity (GR), this article argues that textbooks written during, and shortly before, the so-called Renaissance period constitute a unique record of the intellectual forces that shaped the field of general relativity and gravitation as we know it today. They do so in particular because taken together, they offer a broad, if not an almost complete, variety of stories that the most influential figures of this re-invention period held about what GR should fundamentally be about.

The term Renaissance of General Relativity refers to an historical period that essentially began  in 1955 with the first international conference on general relativity in Bern, Switzerland -celebrating 50 years of relativity- and ended in the early 1970s when the field had established its place in the mainstream of physics. A characteristic feature of the scientific practice of this period was the explicit questioning and examination of the fundamentals of the theory itself. This stood in stark contrast to previous decades, in which general relativity was mainly being used, unaltered in its core, for unified field theory purposes, for application to astronomical theorizing, to quantize the gravitational force within the program of quantum field theory, or within mathematical physics - if it was being considered at all. These aspects were previously pointed out e.g. in \citep{Blum2015a,Blum2016, Eistenstaedt1992}.
Equally importantly, the Renaissance entailed a process of institutional consolidation, particularly through the worldwide establishment of international research centers, as has most thoroughly been worked out by Roberto Lalli in \citep{Lalli2017}. These research centers were all headed by at least one senior scholar, the principal investigator (PI), in a position to attract postdocs and to supervise PhD students. In this way, respective research agendas were directly passed on to younger generations, thus influencing the course of the field's development. 

Most crucial for this paper's historiographical argument --that the textbooks on GR during that period encapsulate a significal part of the intellctual forces that shaped the field at that time-- is the finding that almost every Renaissance PI published at least one textbook narrating the core of their respective research program.\footnote{The basic defining feature for a book to be included here as a textbook is its primary ambition to introduce the reader to (specific topics of) general relativity; be it at a beginner's or at an advanced level. Most books included here, in fact, have the intention to be a self-contained introductory course on the matter. However, for the sake of the historiographical argument, also a monography on a more advanced topic and even an edited book containing chapters by several authors was included. The reasons for these exceptions will be explained at the corresponding parts of the text.}
These Renaissance textbooks thus bundle and condense the main research agendas of the period, i.e. the ways the most influential figures pictured what GR was, or should be, about. A close examination, in all technical detail, of the different stories about GR contained in these books can therefore provide valuable and rich evidence of how GR was negotiated and reinvented at the conceptual level during this period. 

If textbooks are further regarded as deeply embedded in their local, broader historical, and the author's personal contexts, they can additionally be analyzed not only as products of those contexts and articulations of their specific aspects but also as agents that, in turn, reshape them. A comprehensive analysis of the distinct stories these textbooks tell about GR maytherefore be integrated with an examination of: (A) how each textbook manifests its various contexts, including its author’s research trajectory, personal interests, and institutional influences; and (B) the ways in which each textbook generates impact, for example, by disseminating research agendas or attracting younger scholars to these topics. Such analyses are particularly valuable for clarifying the role these textbooks played within the Renaissance period.

Textbook scholars have long emphasized the value of textbooks for understanding the history of science (for overview studies, see, e.g., \citep{Simon2016,Simon2013,Badino2013a}), criticizing in particular the widespread belief that research and teaching are fundamentally distinct; a belief that is especially persistent because textbooks are often regarded as formats that merely display the standard knowledge of their time. The Renaissance textbooks mentioned here offer plentyful opportunities to challenge this belief.  These textbooks do not merely recite established knowledge about GR; rather, they present the subject in a way that reflects the author’s own research program and their critique of the fundaments of the theory, which often includes the introduction of new objects and methods. Beginning with the most basic notions, they guide the reader up to the forefront of current research. Consequently, any student learning GR from such a textbook is inevitably shaped by it in their understanding of what the theory’s basic constituents are, what counts as its correct interpretation, why and how the theory is useful, which methods are appropriate for its application, and which research questions are legitimate. Not only does this manifest a rapidly changing, and therefore incomplete, field, but also it's inherent locality.

There has been previous work examining the history of physics, and that of GR in particular, through textbooks.\footnote{I thank an anonymous reviewer for drawing my attention to the rich literature on textbooks in the history of chemistry and other fields, such as \citep{Lundgren2000}, \citep{BensaudeVincent1990}, \citep{GarciaBelmar2006} (chemistry), \citep{Simon2011} (physics), \citep{Smyth2001} (psychology), \citep{BensaudeVincent2006}, \citep{Myers1992}, \citep{Vicedo2012}, \citep{BertomeuSanchez2006} (more general accounts).} David Kaiser, for example, uses textbooks, along with lecture notes, exercise sets, exam problems, and other classroom teaching techniques, to examine how students were instructed to practice physics. In doing so, he addresses a distinctly Kuhnian question: how the training of young physicists shaped their later professional activity and thus the development of the entire discipline. 
In this sense, Kaiser's edited volume "Pedagogy and the Practice of Science: Historical and Contemporary Perspectives" \citep{Kaiser2005b}, considers the "constitutive roles played by pedagogy in making modern science and engineering happen," \citep[p.7]{Kaiser2005b} at various instances in history and in various local contexts. Thereby emphasizing in particular the historicity of scope, content and pedagogical aspects of the textbook.

In his earlier "A $\Psi$ is just a $\Psi$? Pedagogy, Practice, and the Reconstitution of General Relativity, 1942-1975" \citep{Kaiser1998}, Kaiser found that "[b]y studying how various textbooks presented the subject of general relativity, and what kinds of problems these texts asked students to calculate, we can follow a transition from a predominantly geometrical conception of gravitation to a more dynamical one" \citep[p.323]{Kaiser1998}. As this present paper introduces significantly more textbooks than Kaiser in his paper, which in fact only studies Peter Bergmann's 1942 "Introduction to the Theory of Relativity" \citep{Bergmann1942} in detail (plus lecture notes by Richard Feynman and problem sets by Sidney Coleman), it can be concluded here, that there was not only a transition from a predominantly geometrical conception of gravity to a more dynamical one in this period, but in fact to a multiplicity of conceptions.

In "A Tale of Two Textbooks: Experiments in Genre" \citep{Kaiser2012}, Kaiser analyzes two case studies, Misner, Thorne and Wheeler's "Graviation" (1973) \citep{Misner1973} and Fritjof Capra’s "The Tao of Physics" (1975), to examine the question of what constitutes a textbook and the concrete ways in which this category manifests. This exercise was repeated by Florian Laguens in \citep{Laguens2025} for the cases of Arthur Eddington's books \citep{Eddington1920} and \citep{Eddington1923}.\footnote{I thank Gal Ben Porath for making me aware of Florian Laguens article.} Laguens adds to this analysis the dimension that textbooks may also have a philosophical scope.

Finally, in his "Drawing Theories Apart" \citep{Kaiser2005a}, Kaiser examines how knowledge and usage of Feynman diagrams spread, initially through informal mentoring and network building, and later, between 1955 and 1970 --a period that coincides with the Renaissance of GR-- also via scientific textbooks (see especially the book's Chapter 7). Regarding the nature of the respective textbooks, Kaiser finds that "[i]f we are looking for finished theories or the logical analysis of theories’ structures, one place we will not find them is in the diagrammatic textbooks published between 1955 and 1970" \citep[p.255]{Kaiser2005a}. "Instead," Kaiser explains, "textbook authors worked hard to present just those “productive methods” that they believed graduate students, and even their own colleagues, needed to practice in order to deploy them confidently in their research" \citep[p.254]{Kaiser2005a}. %\footnote{The work by Kaiser on the topic of scientific textbooks is not exhaustively quoted here. There is, for example, also his work with Warwick [...] that is often quoted in the historical literature on textbooks.}

Aside from David Kaiser's work, there is, for example, Massimiliano Badino and Jaume Navarro's edited volume "Research and Pedagogy: A History of Quantum Physics through Its Textbooks" \citep{Badino2013}, which also challenges this Kuhnian view that scientific textbooks only appear during times of normal science and present a completed and neutral body of knowledge to new learners. In its focus on early textbooks on quantum theory the volume aims at "de-black-boxing the process of construction of truth in textbooks during a period of crisis" \citep[p.15]{Badino2013a} and "contribute to the study of textbooks as agents of research" \citep[p.16]{Badino2013a}. In some of the case studies presented in that volume a similar conclusion to one of the main claims of this present paper is made: that the textbooks were used as means to consodilate their own specific research program and their own, individual way to view the new developments within, in the case of Badino and Navarro's volume, quantum theory. This opens, as the editors remark in their introduction to the volume, the issue of the de-localization of scientific knowledge. An aspect that is, as mentioned above, for the same reason also very apparent in the multiple Renaissance research centers spread across the globe.\footnote{A further reference that may well be mentioned here is  Gunter Lind's book "Physik im Lehrbuch, 1700-1850. Zur Geschichte der Physik und ihrer Didaktik in Deutschland" (1992) \citep{Lind1992}. This work understands itself as a contribution to the history of didactics, examining "the influence of the academic discipline on the goals, content, and methods of instruction" ("Einfluß der Fachwissenschaft auf Ziele, Inhalte und Methoden des Unterrichts") \citep[p.1]{Lind1992},  including several texbook case studies. Which, in a sense, represents the opposite ambition to the one presented here.}

The primary purpose of this paper is to elaborate on the point that each Renaissance PI produced a textbook presenting the story of GR in their own way, and, on this basis, to argue for the significant historical value of these textbooks as tools for studying the Renaissance period.
To this end, the paper is structured as follows. Section \ref{s.textbooks} first gives a short history of GR textbooks, treating the first generations of GR textbooks, from Hermann Weyl's seminal "Raum, Zeit, Materie" (1918) to those published in the years immediately preceding the Bern conference of 1955. The established periodization of work in general relativity becomes visible in the way these textbooks treat the subject. However, analysis of the textbooks suggests that interest in GR had already begun to grow and to change in character in the years between the Second World War and the Bern conference.
Section \ref{s.Renaissance} then provides an overview of the textbooks authored by the first principal investigators of the Renaissance research centers as defined, e.g., in \citep{Lalli2017}, accompanied by brief characterizations of the corresponding narratives of GR. These stories bear witness to where the field stood at the beginning and throughout its reinvention: to how GR was being perceived as a theory, to tensions over interpretation between competing schools, and to research directions that, as we now know, would profoundly reshape the field.
The final section, \ref{s.nextgen}, offers a glimpse of the wide range of GR textbooks published in the 1960s, 1970s, and 1980s. including those of later Renaissance PIs. Without being able to go into detail for each textbook, it highlights several characteristic features of what the field had become by then.

The textbook story characterizations in Section \ref{s.Renaissance} are deliberately brief: they provide just enough of each narrative of GR to give a sense of its content and to make visible how different the authors' research programs were at that time. A detailed investigation of each textbook consciously remains to be done. And this is the second aim of the paper, as it serves as an open invitation to historians to undertake such studies.  
The ways in which these Renaissance textbooks can be studied are of course plentiful and leave room for the interests and creativity of the respective historians. This paper emphasizes the dimension of the story each book tells about GR, valuing especially technical and conceptual aspects, because these stories reveal the different conceptions of the theory during the Renaissance period and thus provide an overview of the intellectual forces that shaped the field, as mentioned at the beginning. They can further be read and studied as manifesting their respective local, national and institutional contexts, and as acting back on those contexts through the book itself.

Beyond the stories narrated and the conceptual aspects, there are further dimensions of the textbook worth studying for the historical understanding of the Renaissance. These include pedagogical practices, the materiality of the book, and the role of the publisher and the situation of the publishing industry at the time. As mentioned before, the global character of the Renaissance period invites special attention to how differing local or national textbook and research cultures become visible in these books and distinguish them from one another, and how the dissemination of a textbook within other research schools -and the potential appropriation of its story of GR— shaped the formulation of research questions elsewhere. Beyond the individual volume, the textbooks thus also deserve to be studied in subsets or collectively, mapping and comparing the different directions in which the Renaissance PIs aimed to take the theory.

%----------------------------------------------------------------------------------------
%	Content
%----------------------------------------------------------------------------------------
\section{First Generations of GR Textbooks}\label{s.textbooks}

The Renaissance of General Relativity refers to the period from roughly 1955 to the early 1970's, during which research on general relativity experienced significant conceptual advancements and institutional growth, in particular the consolidation of a research community. The institutional growth came in the shape of the establishment of several research centers mentioned before, each typically led by a senior principal investigator (PI) guiding a group of postdoctoral researchers and doctoral students, as is well documented in the work of Roberto Lalli, Jürgen Renn, and Alexander Blum \citep{Lalli2017,Blum2015a,Blum2016}. These centers also acted as intellectual schools, where knowledge of general relativity was passed on to students and junior researchers. One of the emphases of this article is that this knowledge was channeled through the lens of the PI’s particular research interest and thus had a bit of a different spin depending on the respective school or center. This becomes manifest in the school-specific textbooks that were published during, or shortly before, the renaissance period. Giving an overview over these textbooks is the focus of the following sections. 
Leading up to this overview, a brief historical contextualization of relativity textbooks is given.

Some of the most influential textbooks on GR were written and published prior to the Renaissance, notably
\begin{enumerate}
	\item Hermann Weyl, "Raum - Zeit - Materie" (1918) \citep{Weyl1918a}
	\item Wolfgang Pauli, "Relativitätstheorie" (1921) \citep{Pauli1921}
	\item Arthur Eddington, "The Mathematical Theory of Relativity" (1923) \citep{Eddington1923}.
\end{enumerate}
They belong to a first wave of GR books that were published within a few years after Einstein came forward with the theory. According to Arnold Sommerfeld, at that time there was an "apparently insatiable demand for explanations of the theory of relativity, both popular and highly scientific, which is particularly prevalent in Germany" \citep[p.III]{Pauli1921}. 

Accordingly, more textbooks than these three probably most famous ones were written then. For example, August Kopff's "Die Grundzüge der Einsteinschen Relativitätstheorie" (1921) \citep{Kopff1921}, Max von Laue's "Die Relativitätstheorie - Zweiter Band: Die allgemeine Relativitätstheorie und Einsteins Lehre von der Schwerkraft" (1921) \citep{Laue1921} and Einstein's own "The Meaning of Relativity: Four Lectures Delivered at Princeton University, May 1921 by Albert Einstein" (1922) \citep{Einstein1922}. And in America, there was e.g. Manuel Amoroso Costa's "Introdução à Teoria da Relatividade" (1922) in Brazil\footnote{Thiago Hartz et al mention this textbook in their historiographical overviewpaper \citep{Hartz2025} (in Portuguese). Jean Eisenstaedt and Júlio C. Fabris wrote about this textbook in their brief paper "Amoroso Costa e o primeiro livro brasileiro sobre a Relatividade Geral  (Amoroso Costa and the first Brazilian book on General Relativity)" \citep{Eisenstaedt2004}.} \citep{Costa1922} and George David Birkhoff and Rudolph Ernest Langer's "Relativity and Modern Physics" (1923) in Harvard, U.S.A \citep{Birkhoff1923}.
 More qualitative presentations were published as well, directed to less specialized or even lay audiences, like Erwin Finlay Freundlich's "Die Grundlagen der Einsteinschen Gravitationstheorie" (1916) \citep{Freundlich1916}, Eddington's "Space, Time and Gravitation" (1920)\footnote{The paper \citep{Laguens2025} explains in detail the differences of Eddington's two textbooks.} \citep{Eddington1920}, Max Born's "Die Relativitätstheorie Einsteins und ihre physikalischen Grundlagen" (1920) \citep{Born1920}, Hans Thirring's "Die Idee der Relativitätstheorie" (1921) \citep{Thirring1921} and of course Einstein's own "Über die spezielle und die allgemeine Relativitätstheorie (gemeinverständlich)" (1916) \citep{Einstein1916b}.\footnote{The topic was also quickly picked up for philosophy textbooks, such as Moritz Schlick's "Raum und Zeit in der gegenwärtigen Physik, zur Einführung in das Verständnis der allgemeinen Relativitatstheorie" (1917) \citep{Schlick1917}, Hans Reichenbach's "Relativitätstheorie und Erkenntnis a priori" (1920) \citep{Reichenbach1920}, Ernst Cassirer's "Zur Einsteinschen Relativitätstheorie" (1921) \citep{Cassirer1921} and Reichenbach's "Philosophie der Raum-Zeit-Lehre" (1928) \citep{Reichenbach1928}.}

One can roughly recognize the different periods of research on general relativity that the study of its history has found in the pattern of new editions and republications of these textbooks.\footnote{It seems interesting that this pattern is not visible in the philosophy textbooks mentioned in the previous footnote. While Reichenbach's \citep{Reichenbach1920}, \citep{Reichenbach1928} and Cassirer's \citep{Cassirer1921} books did not even see a second edition (nor a reprint before the 2000s), Schlick's book \citep{Schlick1917} saw three new editions until 1922 but none beyond the early twenties and no English translation (and with a new reprint only in 1983). Reichenbach's \citep{Reichenbach1928} was after all published in English in 1958 \citep{Reichenbach1958}, translated by his second wife Maria.}
This relates espsecially to the so-called Low Watermark period of research in GR, beginning the mid 1920s and leading up to the Renaissance, where interest in this topic was diminished. 
While all of the books mentioned above saw several editions or reprints until 1925, with the exception of Amoroso's and Birkhoff and Langer's books, which did not see new editions at all,\footnote{Amoroso's textbook was reprinted in 1995, however.} there were no new ones after this year and before the end of the Second World War. With one notable exception\footnote{Actually, there was one further tiny exception, which was Eddington's "Space, Time and Gravitation" which got one further reprint in 1928.}: 
Einstein's "Über die spezielle und die allgemeine Relativitätstheorie", which saw two new editions in the 30s and three further new editions until including 1945 (including English translations). 
From 1945 onwards, new editions of the other books mentioned slowly started to appear as well: In the second half of the 1940s there was a new (and final) edition of Thirring's book and two new editions of Einstein's "The Meaning of Relativity" (which continues to be newly edited and reprinted until today as well). 

It seems curious that renewed interest in the then "classic" texts on GR arose already before 1955 that is, before the Renaissance of GR "officially" started. Weyl's book was translated to English in 1952 and Pauli's book was supposed to be translated in 1951 \citep[p.IX]{Pauli2000} (however eventually only appeared in English 1958). 
Von Laue's book got a new edition in 1952/53 and Eddington's "Space, Time and Gravitation" in 1952 as well. Only Born's book and Eddington's "The Mathematical Theory of Relativity" were published in new editions later on (1962 and 1975, respectively - while Kopff's and Freundlich's books did not receive renewed interest at all).\footnote{To be precise, it was the English translation of Born's book from 1924 that saw a new, heavily revised, edition in 1964 on the US  American market. A new German edition was published in 1964, based on the previous English one.} 

However, this does not mean that there were no new textbooks that included an account of GR in this intermediate period between 1925-1945, though there seem to have been only few. There was, for example, Oswald Veblen's "Projektive Relativitätstheorie" (1933) \citep{Veblen1933}\footnote{Strictly speaking, this work is not a textbook, but rather a long treatise included as part of a book. It is, however, still included in this list as a) this is the outcome of a research direction involving GR which is truly unique at that time and b) Veblen's interest in GR lies at the very foundation of the Renaissance of GR in the USA.}, Richard Chace Tolman's "Relativity, Thermodynamics, and Cosmology" (1934) \citep{Tolman1934}, Lev Davidovich Landau and Evgenii Mikhailovich Lifshitz' "Теоретическая физика; т. 2: Теория поля" (1941) \citep{Landau1941} (first English translation in 1951 as "The Classical Theory of Fields" \citep{Landau1951}), Peter Gabriel Bergmann's „Introduction to the Theory of Relativity“ (1942) \citep{Bergmann1942} and Otto Heckmann's "Theorien der Kosmologie" (1942) \citep{Heckmann1942}. 

In the following period, until 1955, the year the famous Bern conference took place and at which historians place the start of the Renaissance period, there were for example John Lighton Synge and Alfred Schild's "Tensor Calculus" (1949) \citep{Synge1949}, Cornelius Lanczos' „The Variational Principles of Mechanics" (1949) \citep{Lanczos1949}, George Yuri Rainich's "Mathematics of Relativity" (1950) \citep{Rainich1950}, Erwin Schrödinger's "Space-Time Structure" (1950) \citep{Schroedinger1950}, Christian Møller's „The Theory of Relativity“ (1952) \citep{Moeller1952}, Hermann Bondi's "Cosmology" \citep{Bondi1952} and Pascual Jordan's "Schwerkraft und Weltall" (1952) \citep{Jordan1952}.

The way the story about general relativity is told in some of these newer books, however, differs significantly from the ones stemming from the first decade after its formulation by Einstein. While the early textbooks generally aimed at grappling with and making comprehensive the new physical and mathematical concepts, decades later this was not necessary anmore. The mathematics was much more routinely presented and the general idea of GR was subsumed into (or appropriated by) other stories about physics, or mathematics. 
In Synge and Schild's textbook, GR was boiled down to being an application of tensor calculus, the Einstein tensor being just an example of a tensor to consider. Lanczos treated GR as an application of variational principles, the Einstein equation being the result of yet another well chosen fundamental action. He very tellingly began his book with the folloing quote by Goethe, "Was du ererbt von deinen Vätern hast, Erwirb es, um es zu besitzen." Similarly, Landau and Lifshitz incorporated the relativistic gravitational field as part of a general treatment of (classical) field theory, including usage of variational principles, with a primary focus on all kinds of aspects of the electromagnetic field.\footnote{Karl Hall's "'Think Less about Foundation': A Short Course on the Course of Theoretical Physics of Landau and Lifshitz" \citep{Hall2006} examines how Landau and Lifshit's volumes influence the style of research in physics in the Soviet Union.}

Put differently, these narratives reframed, or rebranded, (aspects of) general relativity to incorporate it as merely a new example, a new development, into structures and concepts already known, playing down the exceptionality of GR. 
Contrasting to this were for example the textbooks by Tolman and Heckmann, who used the theory of general relativity to apply it to thermodynamics and/or cosmology,\footnote{Tolman's very ambitious program for the book was "first to give a coherent and fairly inclusive account of the well-known and generally accepted portions of Einstein's theory of relativity, second to treat the extension of thermodynamics to special and then to general relativity, and third to consider the application both of relativitstic mechanics and relativistic thermodynamics in the construction and interpretation of cosmological models" \citep[p.1]{Tolman1934}.} and were thus pioneers of a new field, relativistic cosmology. 
Also Hermann Bondi's cosmology book  \citep{Bondi1952} falls into this category. His ambition therein is to "present cosmology as a branch of physics in its own right"; he therefore gathered observational evidence and reviewed several cosmological theories, among them relativitstic cosmology, but also his version of the steady-state theory that rather rejected general relativity. 

But there were also new textbooks in this period that dealt with general relativity for its own sake, that carefully developed its basic ideas (hand in hand with the necessary maths) and that were thus very close to the ambition of the first generation of textbooks. Bergmann's, Møller's, Schrödinger's and Rainich's books belong to this class. Being Einstein's coworker at the time, Bergmann's ultimate ambition in his textbook was to be able to treat unified field theory. With this, he directly connects to Weyl\footnote{Weyl had not yet worked out his geometric unified field theory when the first edition of "Raum-Zeit-Materie" \citep{Weyl1918a} came out.}, Pauli and Eddington's work from the early 1920s and their textbooks \cite{Weyl1919a} \citep{Pauli1921} and \citep{Eddington1923}. 

Schrödinger's book and Veblen's treatise should also be understood as being part of this tradition. While Schrödinger's book is a thorough explorative investigation of the notions of general covariance and of affine connection, as well as of their conceptual implications, Veblen's book is the attempt to unify general relativity with electromagnetism by means of projective geometry. However, while sharing the unifying ambition with Schrödinger, Bergmann et al, Veblen's approach is qualitatively different. Instead of searching for a generalization of the affine structure, or something alike, Veblen's approach is to cast the theory in an entirely new language.\footnote{At that time, there was actually a debate within mathematics on whether or not projective geometry is just another affine geometry, or actually something entirely different. Veblen's school opted for the latter option. See e.g. \citep{Lessel2025} for more details.} Jordan, in contrast, is not primarily interested in the unification gravity and electromagnetism, but makes use of a formalism developed to this end, projective relativity, in order to implement his idea of a varying gravitational constant \citep{Jordan1952}.

Rainich and Møller, in turn, were not interested in unified field theory at all in their books. While Rainich gives a straightforward exposition of general relativity's known mathematical and (despite the title) physical aspects, without any obvious agenda of his own, Møller's agenda is to train students in relativity in order to prepare them to tackle relativistic quantum mechanics. He does not aim to alter the theory in the manner of unified field theory, but he chooses a presentation in which "the four-dimensional formulation of the theory plays a less dominant role [...]", since "[...] in a textbook of today I think it is useful to stress again the fundamental physical difference between space and time, which was somewhat concealed by the purely formal four-dimensional representations." Therefore, "[...] a three-dimensional formulation, which gives a better insight into the physical meaning of the theory, is frequently given", leading to a "[...] reintroduction of dynamical concepts into the gravitational theory [...]."

%------------------------------------------------------------------------
\section{A Need for New Textbooks: The Renaissance of General Relativity}\label{s.Renaissance}
%------------------------------------------------------------------------
%
The previous analysis indicates that interest in GR, in one way or another, slowly built up again --and transformed-- already in the years before the Bern conference (1955), that is, before the 'official' beginning of the Renaissance period. Aside from those who downplayed the exceptionality of general relativity and hid it in already known conceptual frameworks, there were those who a) applied it to cosmology, b) investigated and solidified its mathematical foundation, c) sought for generalizations (serving a certain purpose), d) recasted the theory in the mantle of new mathematics, e) re-narrated the story of GR, for example in order to be able to make it accessible to combinations with quantum theory, or performed a combination of these things.\footnote{This categorization is based solely on the previous analysis of GR textbooks. This means that there may have been further ways to treat GR around that time. This categorization, in any case, essentially coincides with Blum, Lalli and Renn's work, which found that during the Low Watermark Period GR was mainly used for unified field theory, cosmology and quantum gravity and as an object of mathematical enquiry, see e.g. \citep{Blum2015a}.}

Some of the authors mentioned in the last section would become principle investigators (PIs) of the previously mentioned "Renaissance centers", such as Jordan and Bergmann. And within a decade or so after the Bern conference, almost all of these centers had brought about their own textbook on general relativity. This latter fact clearly reflects a general demand for (new) textbooks during that time for the training of an increasing number of physics students, amplified by what David Kaiser called the "new age of textbooks", beginning in the 1950s \citep[p.254ff]{Kaiser2005a}. As Kaiser explains, there was a vast expansion in physics publishing after the war, as publishing houses began to recognize a lucrative market for science publishing in the postwar nuclear age and were thus especially eager to publish all kinds of (text)books on topics in physics. The increase of GR textbooks in particular can be seen to align with the already known fact that the overall published research output on GR significantly increased in this period, which is in fact one of the central tenets of the Renaissance periodization, see \citep[Fig. 2.1, 2.2]{Lalli2017}.

This circumstance provided the authors of Renaissance textbooks with the opportunity to tell the story about GR from the perspective of their respective research program. 
The following table provides an overview of school-specific GR textbooks that were written by the initial PIs of the Renaissance research centers as found in \citep{Lalli2017,Blum2015a}.
The horizontal lines thereby separate the different national contexts. Each PI ist considered to represent a different school, even if they were affiliated with the same institution. Within each national context, the stated texbooks are ordered chronologically.
\begin{table}[h]
	\centering
	\begin{tabular}{llll}
		\hline
		Location    & Author              & Textbook                                 & Year\\ 
		\hline
		Dublin      & C. Lanczos          & The Variational Principles of Mechanics  & 1949 \\	               
	              	& E. Schrödinger      & Space-Time Structure                     & 1950 \\
	                & J. L. Synge         & Relativity: The General Theory           & 1960\\
		\hline
		London/     & H. Bondi            & Cosmology                                & 1952 \\
	    Cambridge   & F. Hoyle            & Astronomy and Cosmology: A Modern Course & 1975 \\
	                & P. Dirac            & General Theory of Relativity             & 1975 \\
		\hline
		Hamburg     & O. Heckmann         & Theorien der Kosmologie                  & 1942 \\
	            	& P. Jordan           & Schwerkraft und Weltall                  & 1952 \\
	    Berlin    	& A. Papapetrou\footnote{Part of Fock school, therefore no early textbook?}       & Lectures on General Relativity           & 1975 \\
		\hline
		Copenhagen  & C. Møller           & The Theory of Relativity                 & 1952 \\
		\hline
		Bern        & A. Mercier et al    & On General Relativity                    & 1979 \\
		\hline
		Paris       & A. Lichnerowicz     & Théories relativistes de la gravitation  &      \\
		            &                     & et de l’électromagnétisme                & 1955  \\
	            	& M.-A. Tonnelat      & La théorie du champ unifié d'Einstein    &       \\
	            	&                     &  et quelques-uns de ses développements & 1955 \\
	    \hline
	    Leningrad   & V. Fock             & Теория пространства, времени и тяготения & 1955 \\
	    \hline
	    Warsaw      & L. Infeld \& J. Plebański & Motion and Relativity              & 1960 \\
	    \hline
	    Syracuse    & P. Bergmann         & Introduction to the Theory of Relativity & 1942 \\
	    RIAS        & L. Witten           & Gravitation: An Introduction to          & 1962 \\
	                &                     & Current Research                         &      \\
	    Chapel Hill & B. DeWitt           & Dynamical Theory of Groups and Fields    & 1965 \\
	    Stevens     & J. Anderson         & Principles of Relativity Physics         & 1967 \\
	    Princeton   & J. Wheeler          & Gravitation                              & 1973\\
	    \hline
	\end{tabular}
	\caption{School-specific texbooks on GR during the Renaissance of GR}
\end{table} 

There were a few schools, or early contributers to the Renaissance, that did not bring about a textbook, or another type of publication of comparable format. Among the research centers treated in \citep[p.16]{Lalli2017}, these were 1) Robert Debever and Jules Géhéniau at the Free University of Brussels (Belgium), 2) Oskar Klein at Stockholm University (Sweden), 3) Nathan Rosen at the Technion in Haifa (Israel) and 4) Frederik Belinfante at Purdue University (USA). These cases are interesting, because they show that independent and influential research lines don't necessarily end up showcasing and advertising themselves through a textbook.\footnote{Strictly speaking, one would need to add P.A.M. Dirac to this list. See below.}

While research center 1) mainly consisted of Debever and Géhéniau themselves, focussing on work on invariants of the Riemannian tensor \citep[p.142]{Lalli2017}, research center 2) mainly consisted of Oskar Klein himself\footnote{Klein's only successful PhD student on GR, Bertel Laurent (\citep[p.152]{Lalli2017}), however did publish a textbook on GR, titled "Introduction to Spacetime: A First Course on Relativity". This book, in any case, appeared only in 1995 (with World Scientific) and thus falls out of the scope of this paper.}, focussing on combining quantum theory with general relativity \citep[p.152]{Lalli2017}. Research center 4) also mainly consisted of Belinfante himself, focussing on the quantization of the gravitational field. In contrast to centers 1,2 and 4, research center 3) was, in fact, rather influential in terms of education of new generations of relativists. 
And while Rosen himself did not author a textbook, his student Moshe Carmeli became a prolific textbook writer. His first single-authored book "Group Theory and Relativity" appeared in 1977 \citep{Carmeli1977}. Asher Peres, further, co-edited the volume "Relativity and Gravitation" that appeared in 1971 with Gordon and Breach, based on a seminar held at Technion in July 1969. If one were to study the research environment of Israel around that time, this book would surely provide very useful evidence.\footnote{In 1959, Rosen was joined by Gerald E. Tauber \citep[p.149]{Lalli2017}, who authored and edited several books related to general relativity and gravitation, directed, however, to a young, or lay audience.}

The PIs of the Renaissance research centers that focused on cosmology had essentially already published their school-building treatises prior to the Bern conference. (With the exception of Fred Hoyle, who came forward with his overarching course on cosmology and astronomy, prominently defending steady state theory, in 1975 \citep{Hoyle1975}, that is, towards the end of the Renaissance, and before that only wrote popular scientific accounts, such as \citep{Hoyle1955}.) Their students would, to a large degree, go on to work on GR proper.

The way general relativity is otherwise treated in these Renaissance textbooks in a sense continues along the lines of investigation  a)-e) listed at the beginning of this section. However, with the major difference that there were no new attempts at downplaying and subsuming the theory into other, already known formal frameworks. Quite on the contrary: one can clearly see how the theory itself is being worked with. This lead to entirely new modes of narration that a) included the introduction and application of new mathematical objects and methods (and with this, perspectives) to the theory, or b) involves serious and explicit attempts to flesh out observable quantities from the theory. 

In Paris, both Lichnerowicz' and Tonnelat's groups grappled with unified field theories. However, in different ways. André Lichnerowicz' specialty was the Cauchy (initial value) problem in GR\footnote{More on the role of the Cauchy problem and the related role of analyticity during the Renaissance of GR can be read in Alexander Blum and Dominic Dold's contribution to this issue.}, his textbook \citep{Lichnerowicz1955} is an analytic perspective on the solvability of the Einstein-Maxwell system, the Jordan-Thiry theory\footnote{Independently, Lichnerowicz' student Thiry developed a theory similar to that of Jordan mentioned before.} and what he called  the Einstein-Schrödinger (unified field) theory, and the properties of certain solutions. With the Cauchy problem, Lichnerowicz introduced a global perspective to GR, as opposed to constructing (local) solutions, within a given coordinate system. 
Marie-Antoinette Tonnelat in turn, in her textbook \citep{Tonnelat1955} (translated to English and expanded in content in 1966 \citep{Tonnelat1966}), attempts to give a synthesis of the state of the art of work on those unified field theories that are based on a generalization of the affine connection. It is interesting to read what Lichnerowicz wrote in his preface to Tonnelat's book: 
\begin{quote}
	Einstein has just passed away, leaving us, alongside so many completed works, an enigmatic theory that scientists regard, like him, with a mixture of suspicion and hope, but which bears the mark of its creator’s fundamental ambition. In order to solve the enigma, whether to dismiss or reinforce the tempting idea, a great deal of work is necessary. Even if, as is likely, this theory does not yield the decisive key to physical fields, the work will still be fruitful, helping us better understand both the resistance of physical fields to geometrization and the ambition of geometrization itself.
\end{quote}
In this spirit, Tonnelat's emphasis in the book is on comparing and evaluating the physical content of the different formulations of (affine) unified field theory.

Although not concerned with unified field theory, John Lighton Synge’s textbook maintains an explicitly geometric perspective \citep{Synge1960}. Following his book on tensor calculus\footnote{And following a book on special relativity, \citep{Synge1956}.} \citep{Synge1949} discussed before, Synge in 1960 put forth a rigorous mathematical treatment of GR. Therein, he again emphasizes tensor calculus, but this time with a clear commitment to understand general relativity as a physical theory of gravity. Synge characterized the spirit of his book as being both "ironic" and "geometric". Ironic, because "[t]he lust for calculation must be tempered by periods of inaction, in which the mechanism is completely unscrewed and then put together again" \citep[p.VII]{Synge1960}. And the way Synge put the mechanism together again is by systematically utilizing a certain 'world-function', previously introduced by Harold Stanley Ruse, being essentially the square of the geodesic distance (first mentioned on p.VIII). 

Synge's book is "geometric", because "the Riemann tensor, [...] \emph{is} the gravitational field [...]. [...] if I break my neck by falling off a cliff, my death is not to be blamed on the force of gravity (what does not exist is necessarily guitless), but on the fact that I did not maintain the first curvature of my world-line, exchanging its security for a dangerous geodesic" \citep[p.IX]{Synge1960}. Yet the approach is also geometric in its emphasis on the utility of spacetime diagrams and the absolute (again, instead of local) properties tensors have: 
\begin{quote}
	The Principle of Equivalence performed the essential office of midwife at the birth of general relativity, but, as Einstein remarked, the infant would have never have go beyond its long-clothes had it not been for Minkowski's concept. I suggest that the midwife be now buried with appropriate honours and the facts of absolute space-time faced. \citep[p.IX/X]{Synge1960}
\end{quote}

Finally, it is curious how in the beginning of his preface he describes work on general relativity as "splitting hairs in an ivory tower":
\begin{quote}
	Then let the relativist rejoice in the ivory tower where he has peace to seek understanding of Einstein's theory as long as the busy world is satisfied to do its jobs without him. Let him be satisfied with the difficult task of gaining for himself as much understanding as he can, and the still more difficult task of transmitting to others the scraps of understanding he has been able to gain. \citep[p.VII]{Synge1960}
\end{quote}
Yet he concludes this same preface by thanking Felix Pirani for influencing him, through their discussions, to attempt “to turn Riemannian geometry into observational physics (measure the Riemann tensor!)"\footnote{See also Gal Ben Porath's paper in this same issue.} This turn to working out observable features of the theory is, as we already know, characterisitc for the Renaissance of GR and it is striking to see this being manifested so explicitely in this preface. 

Also characteristic is the “unscrewing and putting together” that Synge mentions, which in fact points to an almost philosophical negotiation process concerning what general relativity is actually about—and whether, perhaps, a different formal framework might be more appropriate. Fitting well within this process, yet heading in a different narrative direction, was Paul Adrian Maurice Dirac’s interest in general relativity during that period. In the late 1950s, Dirac put forward a Hamiltonian reformulation of general relativity as a basis for quantizing the theory. His textbook \citep{Dirac1975}, however, was written only after he had retired from Cambridge and moved to Florida. Its content, moreover, is too elementary to bear any trace of the Hamiltonian formulation.

While Lichnerowicz in Paris emphasized the importance of a global perspective in order to be able to discuss the integrability of the field equations, Vladimir Aleksandrovich Fock at Leningrad State University around the same time emphasized a "nonlocal" approach to the theory for similar reasons.\footnote{Fock does not use the precise term "nonlocal" in his textbook, though, where he instead refers to properties of "space as a whole". But he does use this notion, for instance, in \citep[p.326]{Fock1957}. This term, of course, very much reminds of quantum mechanical non-locality and one may wonder if this is a coincidence. Fock was, of course, also very influential within quantum physics ("Fock space", "Hartree-Fock method", ...) and in \citep[p.326]{Fock1957} he meaningfully denotes a matter distribution (a system of masses) on space-time by $\Psi$, where he however does not bother to comment further upon this.} In the introduction to his textbook \citep{Fock1955} he states (quote taken from the English translation \citep[p.18]{Fock1959}): 
\begin{quote}
	In order to construct a theory of gravitation or to apply it to physical problems it is, however, insufficient to study space and time only locally, i.e. in infinitely small regions of space and periods of time. One way or another one must also characterize the properties of space as a whole. [...] This is particularly clear in view of the fact that the equations of the gravitational, or any other field, are partial differential equations, the solutions of which are unique only when initial, boundary or other equivalent conditions are given.
\end{quote}

Instead of pursuing a global approach in the sense of Lichnerowicz to treat this problem, Fock studied spaces which allow for global ('distinguished') coordinates, most prominently so-called harmonic coordinates on asympotitcally flat ('asymptotically homogeneous') spaces. 
At the basis for this, and in fact at the basis for Fock's entire narrative of GR (or rather, of 'theory of gravitation', as he prefers to call it) lies a careful analysis of notions of invariance.\footnote{In his textbook, Fock claims that "[t]he philosophical side of our views on the theory of space, time and gravitation was formed under the influence of the philosophy of dialectical materialism, in particluar under the influence Lenin's "Materialism and Empirocriticism"." This influence is also claimed in \citep{Martinez2020}. However, Fock does not repeat this claim in other accounts of his philosophy, such as in \citep{Fock1957,Fock1966} of GR, but instead, for example, calls it a "consequent logical analysis" in \citep[p.28]{Fock1966}.} Fock fundamentally (and emphatically) distinguishes A) relativity, which manifests itself in the existence of the full Lorentz-group as metric invariance group, and B) general covariance of differential equations with respect to coordinate transformations. As such, the principle of relativity is only applicable in what Fock calls uniform, or homogeneous spacetimes, i.e. spacetimes with constant curvature. In particular, he points out that relativity is in this sense restricted and not generalized on non-uniform spacetimes, which have smaller isometry groups. Due to the principle of equivalence, however, spacetime is at least assumed to be uniform in the infinitesimal, that is, the prinicple of relativity is assumed to hold true infinitesimally in GR, while it does not generally hold for "space as a whole".\footnote{The claim made in \citep[p.119]{Martinez2020}, that Fock emphasized 'the need to consider "space as a whole"' to 'preserve relativity', does not seem quite on point. Instead, Fock's ambition to consider "space as a whole" rather stems from a philosophical and mathematical analysis related to the problem of finding unique solutions to the field equations (see quote given above). Since relativity is still valid in the infinitesimal, it does not need to be preseved, it should just not be confused with the transormations that are undertaken with respect to coordinate transformations. In particular, it seems additionally important to stress that Fock did not doubt the validity of Einstein's gravitational theory, as he explicitely stated in e.g. \citep[p.326]{Fock1957}.} 

A consequence of such a local description is the difficulty to provide meaningful initial or boundary conditions to uniquely determine solutions of the field equation. To this end, Fock aimed to study spaces which accommodate useful properties of uniformity. The simplest of his cases studied is the above mentioned asymptotically homogeneous space, which by definition is uniform at infinity. In this case, all integrals of motion for a system of masses generating local inhomogeneities exist. On these spaces also preferred coordinates exist, the so-called harmonic coordinates, well-defined up to Lorentz transformations, which cover the entire space (i.e. what would nowadays be called global coordinates) and which allow for simpler interpretations of the physical situation at hand.

While Fock ascribes fundamental physical importance to relativity, covariance of equations in contrast is to him "in itself [...] in no way the expression of any kind of physical law" \citep[p.xvi]{Fock1959}\footnote{This should not be misunderstood that Fock is not willing to admit physical significance to expressions which are generally covariant.}, differentiating in particular between reference system and coordinate system. To the effect that Fock advocates the use of 'distinguished' coordinate systems, such as harmonic coordinates, for more intuitive understanding of the physical situation at hand, for simplified calculations and for a solution to the problem of having well-defined initial values. The fact that general covariance as a mathematical property is needed is to him an "almost trivial demand" \citep{Fock1966}.

The judgement of Fock's take on the interpretation of GR in the camp of relativists closeer to Einstein was of rather skeptical, if not dismissive, nature. In the 1965 edition of Born's textbook \citep{Born1920}, the following verdict is given when recommending further literature on the topic in the preface: 
\begin{quote}
	V. FOCK’s book, "Theorie von Raum, Zeit und Gravitation", on the other hand, does present the special theory of relativity correctly and also contains many insightful observations in its more advanced chapters. However, I have reservations about his treatment of the fundamentals of the general theory of relativity. FOCK considers the fundamental insight that led EINSTEIN to the general theory of relativity—the so-called equivalence principle—to be only approximately correct and attempts to circumvent it. This is not only ahistorical but also incorrect [...].
\end{quote}
Where at the end of the quote Born refers to the articles, "Über die Interpretation der allgemeinen Relativitätstheorie von V. Fock" by Roman Sexl (1962) \citep{Sexl1962} and "The Principle of Equivalence" by Fritz Rohrlich (1963) \citep{Rohrlich1963}. Any account of Fock's narrative of GR given in his textbook would thus meaningfully require its reception in the global relativity community. 

Based on his narrative of GR, in any case, one main thing Fock was after was deriving the equations of motion of a system of particles. That is, to study the problem of motion. Coincidentally, this was also the main theme of Leopold Infeld's textbook \citep{Infeld1960} that he co-authored with Jerzy Plebański. Infeld, however, who had already begun working on that problem with Einstein back in the 1930's, approached the matter "more in the spirit of Einstein" \citep[p.7]{Infeld1960}. 
This is, while Fock treated particles of finite size and with nonlocal techniques, Infeld and his group treated particles as singularities of the field within a purely local approach. 
These differing approaches to the problem did also lead to a debate, which is described in \citep{Martinez2020}.
In their book, Infeld and Plebański are very concerned about the concept of point-particles and the gravitational self-fields they may have. In the case most interesting to them, point-particles have finite mass and generate a field which is singular on their world-lines. The book, then, is mainly a story about formally treating this singular behaviour "to find the essential characteristic features of the motion of point particles" \citep[p.15]{Infeld1960} and their interaction with gravity, deploying a highly sophisticated, technical machinery (using approximation methods). Importantly, in contrast to Fock, they at no point question the "logical structure of G.R.T." (which is how they abbreviate the general theory of relativity). 

%--------------------------------------------------
\section{Next Generations of Renaissance Textbooks}\label{s.nextgen}
%--------------------------------------------------
%
In \citep{Kaiser2007,Kaiser2002}, Kaiser observes a correlation, causation even, between growing university class sizes in physics in the U.S.A after the Second World War and applied pedagogical methods in quantum mechanics classes. Acoording to Kaiser, the large amounts of students per class demanded a teaching which was much less philosophical and much more practial, focusing on concrete calculations without in-depth discussions of interpretation. This pattern he sees reproduced in the demands of what textbooks were supposed to provide at the time. In particular, they converged on a specific pedagogical style that included hundreds of (technical) problems and excercises. When enrollment numbers dropped again in the 1970s, some of these exercises turned into interpretative discussion questions again.

Is there a similar development within the field of general relativity? It seems that, perhaps somewhat later than for the case of quantum mechanics, also GR textbooks have increasingly turned into hands-on practical self-teaching devices, rather than the mere development of theory as it was the case in earlier decades. However, this is a question that remains to be explored. Taking a brief glance at Dirac's texbook \citep{Dirac1975} from 1975, in any case, reveals that this book is much more beginner's friendly in it's scope and employed pedagogical strategies, not necessarily designed for the training of future researchers on that topic, than what used to be more customary until the 1960s at least. The overall pace of that book is rather slow, for example developing each crucial concept, like curvilinear coordinates, Christoffel symbols, geodesics and so on, in an individual chapter. There is also no obvious trace of Dirac's own research agenda on Hamiltonian constraints. Instead, this book conveys what Dirac and others may have had conceived as canonical formalistic and interpretative basis of the GR. 

Also Misner, Thorne and Wheeler's textbook \citep{Misner1973} from 1973 is directly adddressed to university physics students, with a very elaborate pedagogical style and unique structre that David Kaiser analyzes in in his paper \citep{Kaiser2012}. This is, as said, remarkably different to the textbook writing of earlier Renaissance PIs and may be explained, as Kuhn may have liked it, with the circumstance that GR has then not only entered the mainstream of physics and part of physics curricula, but has also reached a phase or normal science. 

Going back to the 1960s, some of the Renaissance PIs were just getting started textbook-wise,  and some of those PIs discussed in the previous sections had already put out new editions, or reprints, of their textbooks\footnote{The following list gives the years of all  (as far as was possible to find out) available reprints, newer editions, or translated editions (E: English, D: German) until 1990 of those authors of the above table that have published their first book before 1961.}: Lanczos (1962, 1970, 1986); Schrödinger (1954, 1960, 1985, 1987 (D)); Bondi (1960, 1961); Jordan (1955); Møller (1969, 1972, 1976 (D)); Fock (1959 (E), 1960 (D), 1961,  1964 (E), 1964); Tonnelat (1964 (E)); Bergmann (1976)).\footnote{Lanczos, Schrödinger, Jordan, Lichnerowicz and Bergmann afterwards wrote further textbooks, or monographies, however on different, or only peripherically related topics, which shall not be mentioned here.} 
In a sense, the textbooks whose first edition came out in the 1960s were already authored by the second generation of Renaissance relativists, who put new topics and new emphases on the table. 

Carl Bryce DeWitt was perhaps not a second generation relativist in the sense that he was taught by one of the other Renaissance PIs. However, he was also not one of those PIs who had lived through the Low Watermark period preserving the lore. DeWitt received his PhD in 1950,\footnote{If one would allow to broaden the notion of textbooks somewhat, DeWitt's PhD thesis "On the Theory of Gravitational Interactions, and the Interaction of Gravitation with Light" submitted in 1949 to Harvard University could also count as such.} under the supervision of Julian Schwinger, and from the very beginning focused on one topic: the (canonical) quantization of the gravitational field. His book \citep{DeWitt1965}, published in 1965, is thus a story of (quantum) gravity written from the persepctive of a particle physicist, regarding the gravitational field as a non-Abelian gauge field.\footnote{It was in parts previously published as part of the proceedings of the Les Houches Summer School in 1963, \citep{DeWitt1964}, edited by his wife Cécile DeWitt-Morette.}

James Leroy Anderson was similarly interested in the (covariant) quantization of the gravitational field, having completed a PhD on this topic under the supervision of Bergmann in 1952. In his textbook \citep{Anderson1967} published in 1967, however, there is little trace of this. Whereas DeWitt’s book is more a report on the state of the art of current research, Anderson’s is a textbook in the common sense. His ambition, nonetheless, was “to give an account of relativity physics that differs from the traditional presentations in a number of ways.”
Importantly, both relativity theories are not justified by means of philosophical, a priori reasoning, but instead by supplying, “the justification by relating the consequences of these theories to the physical world” \citep[p.vii]{Anderson1967}. Moreover, this justification is not grounded in experiments long past, but in “the current usages of the theories” \citep[p.viii]{Anderson1967}, such as the Dirac equation and cosmological applications. In particular, this rhetoric sounds far more self-confident than Synge’s apologetic talk of “splitting hairs in an ivory tower” only seven years earlier.\footnote{Integral part of Anderson's presentation in his book is to place "considerable emphasis on the notion of the symmetry group of a theory" \citep[p.vii]{Anderson1967}. In his analysis, however, he does not, as Fock critizied in his book that people wouldn't, distinguish between symmetry in the sense of isometry and symmetry in the sense of covariance.}

Louis Witten became an active researcher in relativity upon joining the Research Institute for Advanced Study (RIAS) in Baltimore, Maryland (USA), perhaps inspired to pursue this new line of research by his preceding postdoctoral stay at Princeton. His book \citep{Witten1962}, published in 1962, is not a textbook in the pedagogical sense, but an edited collection of essays by several authors from various institutions, intended “to describe recent progress in research associated with various features of gravitation as described by the general theory of relativity” \citep[p.v]{Witten1962}. In particular, it is not an account of Witten’s personal research program; it is included in the above list nevertheless, since it arguably demonstrates the increasingly global connectedness of research programmes, which are no longer confined to a local context alone.\footnote{Lalli et al point out in their work how new global mobility, especially by young postocs, decisively influence the development of the field in that period.} 
In some sense, however, this volume has a similar ambition to that of an (advanced) textbook, as it "is aimed at permitting physicists, engineers, and mathematicians [...] to become familiar with the current trends and status of the theory either as part of their scientific background or as a guide to serious study and active participation in the field" \citep[p.v]{Witten1962}. The chapters included treatment of "the present experimental situation regarding gravitation", (exact and approximate) solutions of the field equations, the problem of motion, the Cauchy problem, conservation laws, gravitational radiation, the Hamiltonian form of general relativity, quantization of the gravitational field, treatment of singularities of the field, application to cosmology and, in fact, unified field theory (now in the form of so-called 'geometrodynamics').\footnote{Geometrodynamics is originally due to John Wheeler and contains the basic idea that all of physics might be explained in terms to geometrical and topological properties.}

In the following decades, the market was flooded by textbooks authored by later generations of relativists. Mainly for illustration purposes and without being able to give detailed asessments of these books as was done above, a (certainly incomplete) list of (mostly English language) textbooks which appeared before 1990 is given below, grouped by decade. 
Next to the above mentioned feature that GR textbooks seemed to have predominantly developed into self-teaching devices for university students, also the topical emphazes of the textbooks suggest that there is new significant awareness of the observational potential of GR, which in fact is the defining feature of what is considered to be the second phase of the Renaissance. Remarkable is further that even though GR was throughout its history also (and by some, mainly) seen as a mathematical discipline, the amount of new mathematics being employed to study and develop further the formal apparatus of general relativity reaches a new quality in these decades. The textbook \citep{Mercier1979} by André Mercier, Hans-Jürgen Treder and Wolfgang Yourgrau, published in 1979, and in fact the last book on GR published by one of the "founding PI's" of Renaissance (Mercier), bears witness of of these latter two observations. However, all of these aspects invite further investigations. 

\paragraph{1960's}

Robert Henry Dicke's "The Theoretical Significance of Experimental Relativity" (1964) \citep{Dicke1964}, 
Yvonne Choquet-Bruhat's "Géométrie différentielle et systèmes extérieures" (1968) \citep{ChoquetBruhat1968}, 
Howard Percy Robertson and Thomas W. Noonan's "Relativity and Cosmology" (1968) \citep{Robertson1968}, 
Dennis Sciama's, "The Physical Foundations of General Relativity" (1969) \citep{Sciama1969}, Wolfgang Rindler's "Essential Relativity: Special, General, and Cosmological" \citep{Rindler1969}.

\paragraph{1970's}
Robert Dicke's "Gravitation and the Universe" (1970) \citep{Dicke1970}, James Peeble's "Physical Cosmology" (1971) \citep{Peebles1971}, 
Yakov Borisovič Zel'dovič and Igor Dmitriyevič Novikov's "Relativistic Astrophysics Vol. 1: Stars and Relativity" (1971) \citep{Zeldovich1971}, 
Steven Weinberg's "Gravitation and Cosmology: Principles and Applications of the General Theory of Relativity" (1972) \citep{Weinberg1972},
Roger Penrose's "Techniques in Differential Topology in Relativity" (1972) \citep{Penrose1972}, 
Stephen William Hawking and George Francis Rayner Ellis' "The Large Scale Structure of Space-Time" (1973) \citep{Hawking1973} %(R)
, Moshe Carmeli's "Group Theory and General Relativity: Representations of the Lorentz Group" %(R) 
(1977) \citep{Carmeli1977}.

\paragraph{1980's} 
Hans-Jürgen Treder et al's "Fundamental Principles of General Relativity Theories: Local and Global Aspects of Gravitation and Cosmology" (1980) \citep{Treder1980}, Subrahmanyan Chandrasekhar's "The Mathematical Theory of Black Holes" (1983) \citep{Chandrasekhar1983},
Norbert Straumann's "Allgemeine Relativitätstheorie und relativistische Astrophysik" (1981) \citep{Straumann1981}\footnote{Straumann wrote in his preface: "I hope I have made a small contribution to increasing the study and teaching of the general theory of relativity in German-speaking countries once again. It is, after all, quite absurd that a trained physicist in the second half of the 20th century knows no more about this magnificent theory and its fascinating applications than a cat knows about the Lord’s Prayer." (Translated to Englisch from the German original.)},
Yakov Borisovič Zel'dovič and Igor Dmitriyevič Novikov's "Relativistic Astrophysics Vol. 2: The Structure and Evolution of the Universe" (1983) \citep{Zeldovich1983}, 
Robert Manuel Wald's "General Relativity" (1984) \citep{Wald1984}, 
Roger Penrose and Wolfgang Rindler's "Spinors and Space-Time" (Vol. 1: 1984, Vol. 2:1986) \citep{Penrose1984}, \citep{Penrose1986}, 
 \citep{Treder1980} %(R)
, Bernard Schütz' "A First Course in General Relativity" (1985) \citep{Schutz1985}, Felix Pirani and Mike Crampin's "Applicable Differential Geometry" (1986) \citep{Pirani1986}, 
 Gregory L. Naber's "Spacetime and Singularities: An Introduction" (1989) \citep{Naber1989}, 
Yvonne Choquet-Bruhat's "Graded Bundles and Supermanifolds" (1989) \citep{ChoquetBruhat1989}.

\section{Final Remarks}
%-----------------------
%
The primary intention of this article, as mentioned in the introduction, was to state and elaborate on the observation that almost all Renaissance PIs eventually brought about their own GR textbook, one that detailed and narrated their specific perspective on the topic. The short characterizations of the various stories these textbooks tell about GR were embedded in a brief historical account on GR textbooks, beginning from the conception of general relativity in 1915/16 until, in some sense, the late 1980s. This brief historical analysis of the history of GR textbooks displays both new and known features of the history of GR itself. It confirms the established periodization into a first peak of interest in GR until the mid 1920s, followed by a Low Watermark period, then by the Renaissance period and finally by a consolidation of the field as part of the mainstream of physics. The textbooks of the Renaissance and those written shortly before presented here display the fundamental level on which new and old relativists were re-negotiating the theory. They also provide a first overview over the manifold ways in which GR was re-narrated during the Renaissance period. Remarkably, a brief historical overview of GR textbooks such as this already shows that interest in GR, also on the conceptual level, slowly built up again already in the years preceeding the "official" beginning of the Renaissance period in 1955. This transition period, between the Low Watermark and the Renaissace thus deserves further attention by historians of physics.

This brief analysis of Renaissance textbooks further adds interesting aspects to the question whether or not GR was already in a state of normal science during the Renaissance period. For the creation of the theory, which may be considered as the main revolutionary act, had long passed, but the Renaissance's defining feature is the fundamental re-working and in part also questioning of the theory, which is not characteristic for normal science. Jürgen Renn et al have called this process "the long relativity revolution", as described e.g. in \citep{Blum_2018}. 
According to Thomas Kuhn, however, textbooks only appear after those kinds of debates have settled and a standard theoretical body has formed \citep[Section XII]{Kuhn1962}. While Renaissance textbooks, did rely on some sort of common core that is attributed to the theory of general relativity, the several ways in which the theory has been attempted to be reworked by the authors, for example via the introduction of new methods and objects, the casting of it into entirely new mathematical formalisms, or the ambition to interpet it in entirely new terms, don't allow for those textbooks to be characterized as treating (the same) standard and finished knowledge about GR. The theory was still in the process of being developed and the texbook authors presented here take the reader all the way from what they considered to be the basics of the theory, to the forefront of their current research activity.\footnote{In particular, the textbook authors mentioned here were no pedagogues in the strict sense, they were primarily researchers who took the opportunity to present their view on the field to propective students (and to the world more generally).}
Interestingly though, it seems that as soon as the consolidation of the field as part of the mainstream of physics was accomplished, textbooks did become much more homogeneous and in fact did present a canoncal theoretical body to prospective students. This, however, still needs to be investigated in detail as well. 

This leads to the second aim of the paper, which is to invite fellow historians to study GR textbooks. This invitation is grounded in the paper’s argument that GR textbooks are useful analytic tools for historical understanding, particularly regarding the Renaissance period. The possible modes of analysis by way of textbooks are thereby manifold, as outlined in the introduction. Yet they all share the potential to bridge and fruitfully combine textbook studies with the history of science and, more specifically, the history of intellectual developments within physics, addressing also pressing questions such as local vs. global aspects of scientific advancement.

\section*{Acknowledgements}
The idea of systematically studying Renaissance textbooks arose while preparing a talk for the workshop “Mathematics in the Making,” held at the University of Bonn in May 2025 on the role of lecture notes in the advancement of mathematics. I am thus grateful to the organizers for the invitation. I further greatly thank Ivã Gurgel for many helpful comments and references.
This work was funded by the European Research Council (ERC) under the European Union’s Horizon 2020 research and innovation programme (grant agreement n◦101088528, COGY). I thank the Lichtenberg Group for History and Philosophy of Physics of the University of Bonn for discussions on the topic.
%----------------------------------------------------------------------------------------
%	BIBLIOGRAPHY
%----------------------------------------------------------------------------------------

\renewcommand{\refname}{\spacedlowsmallcaps{References}} % For modifying the bibliography heading

\bibliographystyle{unsrt}

\bibliography{C:/Users/Admin/Documents/Archives/Bibliothek/Bibliothek}

\begin{thebibliography}{100}

\bibitem{Blum2015a}
Alexander Blum, Roberto Lalli, and Jürgen Renn.
\newblock {The Reinvention of General Relativity: A Historiographical Framework
  for Assessing One Hundred Years of Curved Space-time}.
\newblock {\em ISIS}, 106(3), 2015.

\bibitem{Blum2016}
Alexander~S. Blum, Roberto Lalli, and Jürgen Renn.
\newblock {The renaissance of General Relativity: How and why it happened}.
\newblock {\em Annalen der Physik}, 528(5):344--349, May 2016.

\bibitem{Eistenstaedt1992}
Jean Eistenstaedt and A.~J. Kox, editors.
\newblock {\em {Studies in the History of General Relativity}}.
\newblock Birkhäuser, 1992.

\bibitem{Lalli2017}
Roberto Lalli.
\newblock {\em {Building the General Relativity and Gravitation Community
  During the Cold War}}.
\newblock Springer, 2017.

\bibitem{Simon2016}
Josep Simon.
\newblock {\em Textbooks}, chapter {A Companion to the history of science},
  pages 400--413.
\newblock Wiley, 2016.

\bibitem{Simon2013}
Josep Simon.
\newblock {\em {Physics Textbooks and the Textbook Physics in the Nineteenth
  and Twentieth Centuries}}, chapter {The Oxford Handbook of the History of
  Physics}, pages 651--678.
\newblock Oxford University Press, 2013.

\bibitem{Badino2013a}
Massimiliano Badino and Jaume Navarro.
\newblock {\em {Pedagogy and Research. Notes for a Historical Epistemologyof
  Science Education}}.
\newblock Max-Planck-Gesellschaft zur Förderung der Wissenschaften, 2013.

\bibitem{Lundgren2000}
Anders Lundgren and Bernadette Bensaude-Vincent, editors.
\newblock {\em {Communicating Chemistry. Textbooks and Their Audiences,
  1789-1939 }}.
\newblock Science History Publications, 2000.

\bibitem{BensaudeVincent1990}
Bernadette Bensaude-Vincent.
\newblock {A View of the Chemical Revolution through Contemporary Textbooks:
  Lavoisier, Fourcroy and Chaptal}.
\newblock {\em British Journal for the Historyof Science}, 23:435--460, 1990.

\bibitem{GarciaBelmar2006}
Bernadette Bensaude-Vincent Antonio García-Belmar, José R. Bertomeu-Sánchez.
\newblock {\em {The Power of Didactic Writings: French Chemistry Textbooks of
  the Nineteenth Century}}, chapter Pedagogy and the Practice of Science:
  Historical and Contemporary Perspectives., pages 219--251.
\newblock The MIT Press, 2006.

\bibitem{Simon2011}
Josep Simon.
\newblock {\em {Communicating Physics: The Production, Circulation, and
  Appropriation of Ganot’s Textbooks in France and England, 1851-1887}}.
\newblock Pickering and Chatto, 2011.

\bibitem{Smyth2001}
Mary~M. Smyth.
\newblock {Fact Making in Psychology: The Voice of the Introductory Textbooks}.
\newblock {\em Theory and Psychology}, 11:609--636, 2001.

\bibitem{BensaudeVincent2006}
Bernadette Bensaude-Vincent.
\newblock {Textbooks on the Map of Science Studies}.
\newblock {\em Science and Education}, 15:667--670, 2006.

\bibitem{Myers1992}
Gregory~A. Myers.
\newblock {Textbooks and the Sociology of Scientific Knowledge}.
\newblock {\em English for Specific Purposes}, 11:3--17, 1992.

\bibitem{Vicedo2012}
Marga Vicedo.
\newblock {Introduction: The Secret Lives of Textbooks}.
\newblock {\em Isis}, 103(83--87), 2012.

\bibitem{BertomeuSanchez2006}
José~R. Bertomeu-Sánchez, Anders~Lundgren Antonio García-Belmar, and Manolis
  Patiniotis.
\newblock {Introduction: Scientific and Technological Textbooks in the
  EuropeanPeriphery}.
\newblock {\em Science and Education}, 15:657--665, 2006.

\bibitem{Kaiser2005b}
David Kaiser, editor.
\newblock {\em {Pedagogy and the Practice of Science: Historical and
  Contemporary Perspectives}}.
\newblock MIT Press, 2005.

\bibitem{Kaiser1998}
David Kaiser.
\newblock {A $\Psi$ is just a $\Psi$? Pedagogy, Practice, and the
  Reconstitutionof General Relativity, 1942–1975}.
\newblock {\em Stud. Hist. Phil. Mod. Phys.,}, 29(3):321--338,, 1998.

\bibitem{Bergmann1942}
Peter~Gabriel Bergmann.
\newblock {\em {Introduction to the Theory of Relativity}}.
\newblock Prentice-Hall, Inc, 1942.

\bibitem{Kaiser2012}
David Kaiser.
\newblock {A Tale of Two Textbooks: Experiments in Genre}.
\newblock {\em Isis}, 103(1):126--138, March 2012.

\bibitem{Misner1973}
Charles~W. Misner, Kip~S. Thorne, and John~Archibald Wheeler.
\newblock {\em Gravitation}.
\newblock W.H. Freeman and Company, 1973.

\bibitem{Laguens2025}
Florian Laguens.
\newblock {Eddington, Teacher of General Relativity: Another Tale of Two
  Textbooks}.
\newblock {\em Studies in History and Philosophy of Science}, 112:47--56,
  August 2025.

\bibitem{Eddington1920}
Arthur Eddington.
\newblock {\em {Space, Time and Gravitation}}.
\newblock Cambridge University Press, 1920.

\bibitem{Eddington1923}
Arthur Eddington.
\newblock {\em {The Mathematical Theory of Relativity}}.
\newblock Cambridge University Press, 1923.

\bibitem{Kaiser2005a}
David Kaiser.
\newblock {\em {Drawing Theories Apart - The Dispersion of Feynman Diagrams in
  Postwar Physics}}.
\newblock The University of Chicago Press, 2005.

\bibitem{Badino2013}
Massimiliano Badino and Jaume Navarro.
\newblock {\em {Research and Pedagogy:A History of Quantum Physics through Its
  Textbooks}}.
\newblock Max-Planck-Gesellschaft zur Förderung der Wissenschaften, Berlin,
  2013.

\bibitem{Lind1992}
Gunter Lind.
\newblock {\em {Physik im Lehrbuch, 1700-1850. Zur Geschichte der Physik und
  ihrer Didaktik in Deutschland}}.
\newblock Springer, 1992.

\bibitem{Weyl1918a}
Hermann Weyl.
\newblock {\em {Raum, Zeit, Materie}}.
\newblock Springer, 1918.

\bibitem{Pauli1921}
Wolfgang Pauli.
\newblock {\em Relativitätstheorie}.
\newblock Vieweg+Teubner Verlag, 1921.

\bibitem{Kopff1921}
August Kopff.
\newblock {\em {Grundzüge der Einsteinschen Relativitätstheorie}}.
\newblock S. Hirzel, 1921.

\bibitem{Laue1921}
Max von Laue.
\newblock {\em {Die Relativitätstheorie - Zweiter Band: Die allgemeine
  Relativitätstheorie und Einsteins Lehre von der Schwerkraft}}.
\newblock Vieweg, 1921.

\bibitem{Einstein1922}
Albert Einstein.
\newblock {\em {The Meaning of Relativity: Four Lectures Delivered at Princeton
  University, May 1921 by Albert Einstein}}.
\newblock Methuen Co. Ltd., London, 1922.

\bibitem{Hartz2025}
Juliana~Correia Ferreira, Juliana Youssef, and Thiago Hartz.
\newblock O renascimento da relatividade geral: teses e perspectivas
  historiográficas.
\newblock {\em Cadernos de Astronomia}, 6(1):95--112, April 2025.

\bibitem{Eisenstaedt2004}
Jean Eisenstaedt and Júlio~C. Fabris.
\newblock {Amoroso Costa e o primeiro livro brasileiro sobre a Relatividade
  Geral (Amoroso Costa and the first Brazilian book on General Relativity)}.
\newblock {\em Revista Brasileira de Ensino de Física}, 26(2):185--192, 2004.

\bibitem{Costa1922}
Manuel~Amoroso Costa.
\newblock {\em {Introdução à Teoria da Relatividade}}.
\newblock Süsskind de Mendonça \& Cia., 1922.

\bibitem{Birkhoff1923}
George~David Birkhoff and Rudolph~Ernest Langer.
\newblock {\em {Relativity and Modern Physics}}.
\newblock Harvard University Press, Cambridge, 1923.

\bibitem{Freundlich1916}
Erwin~Finlay Freundlich.
\newblock {\em {Die Grundlagen der Einsteinschen Gravitationstheorie}}.
\newblock Springer, 1 edition, 1916.

\bibitem{Born1920}
Max Born.
\newblock {\em {Die Relativitätstheorie Einsteins}}.
\newblock Springer, 1 edition, 1920.

\bibitem{Thirring1921}
Hans Thirring.
\newblock {\em {Die Idee der Relativitätstheorie}}.
\newblock Springer, 1921.

\bibitem{Einstein1916b}
Albert Einstein.
\newblock {\em {Über die spezielle und die allgemeine Relativitätstheorie
  (gemeinverständlich)}}.
\newblock Vieweg, 1 edition, 1916.

\bibitem{Schlick1917}
Moritz Schlick.
\newblock {\em {Raum und Zeit in der gegenwärtigen Physik, zur Einführung in
  das Verständnis der allgemeinen Relativitatstheorie}}.
\newblock Springer, 1917.

\bibitem{Reichenbach1920}
Hans Reichenbach.
\newblock {\em {Relativitätstheorie und Erkenntnis a priori}}.
\newblock Springer, 1920.

\bibitem{Cassirer1921}
Ernst Cassirer.
\newblock {\em {Zur Einsteinschen Relativitätstheorie}}.
\newblock Bruno Cassirer Verlag, 1921.

\bibitem{Reichenbach1928}
Hans Reichenbach.
\newblock {\em {Philosophie der Raum-Zeit-Lehre}}.
\newblock Walter de Gruyter, Berlin/Leipzig, first edition, 1928.

\bibitem{Reichenbach1958}
Hans Reichenbach.
\newblock {\em {The Philosophy of Space and Time}}.
\newblock Dover, 1958.

\bibitem{Pauli2000}
Wolfgang Pauli.
\newblock {\em {Relativitätstheorie}}.
\newblock Springer, 2000.

\bibitem{Veblen1933}
Oswald Veblen.
\newblock {\em {Projektive Relativitätstheorie}}.
\newblock Number~2 in Ergebnisse der Mathematik und ihrer Grenzgebiete.
  Springer, 1933.

\bibitem{Tolman1934}
Richard~C. Tolman.
\newblock {\em {Relativity, Thermodynamics and cosmology}}.
\newblock Oxford University Press, 1934.

\bibitem{Landau1941}
Lev~Davidovich Landau and Evgenii~Mikhailovich Lifshitz.
\newblock {\em {Теоретическая физика; т. 2: Теория
  поля}}.
\newblock Гостехиздат, 1941.

\bibitem{Landau1951}
Lev~Davidovich Landau and Evgenii~Mikhailovich Lifshitz.
\newblock {\em {The Classical Theory of Fields, (Course of Theoretical Physics,
  Volume 2) }}.
\newblock Pergamon Press/Addison-Wesley, 1951.

\bibitem{Heckmann1942}
Otto Heckmann.
\newblock {\em {Theorien der Kosmologie}}.
\newblock Grundlagen der physikalischen Weltbilder. Springer, 1942.

\bibitem{Synge1949}
John~Lighton Synge and Alfred Schild.
\newblock {\em {Tensor Calculus}}.
\newblock University of Toronto Press, 1949.

\bibitem{Lanczos1949}
Cornelius Lanczos.
\newblock {\em {The Variational Principles of Mechanics}}.
\newblock University of Toronto Press, 1949.

\bibitem{Rainich1950}
George~Yuri Rainich.
\newblock {\em {Mathematics of Relativity}}.
\newblock John Wiley \& Sons, 1950.

\bibitem{Schroedinger1950}
Erwin Schrödinger.
\newblock {\em {Space-Time Structure}}.
\newblock Cambridge University Press, 1950.

\bibitem{Moeller1952}
Christian Møller.
\newblock {\em {The Theory of Relativity}}.
\newblock International Series of Monographs on Physics. Clarendon Press, 1952.

\bibitem{Bondi1952}
Hermann Bondi.
\newblock {\em {Cosmology}}.
\newblock Cambridge University Press, 1952.

\bibitem{Jordan1952}
Pascual Jordan.
\newblock {\em {Schwerkraft und Weltall}}.
\newblock Vieweg, 1952.

\bibitem{Hall2006}
Karl Hall.
\newblock {\em {'Think Less about Foundation': A Short Course on the Course of
  Theoretical Physics of Landau and Lifshitz}}, chapter {Pedagogy and the
  Practice of Science: Historical and Contemporary Perspectives}, pages
  253--286.
\newblock MIT Press, 2006.

\bibitem{Weyl1919a}
Hermann Weyl.
\newblock {\em {Raum, Zeit, Materie}}.
\newblock Springer, 1919.

\bibitem{Lessel2025}
Bernadette Lessel.
\newblock {Pascual Jordan's "Erweiterte Gravitationstheorie" - A Historical
  Analysis of its Mathematical Framework}, 2025.

\bibitem{Carmeli1977}
Moshe Carmeli.
\newblock {\em {Group Theory and General Relativity: Representations of the
  Lorentz Group}}.
\newblock McGraw-Hill, 1977.

\bibitem{Hoyle1975}
Fred Hoyle.
\newblock {\em {Astronomy and Cosmology: A Modern Course}}.
\newblock W.H. Freeman, 1975.

\bibitem{Hoyle1955}
Fred Hoyle.
\newblock {\em {Frontiers of Astronomy}}.
\newblock Heinemann Educational Books, 1955.

\bibitem{Lichnerowicz1955}
André Lichnerowicz.
\newblock {\em {Théories relativistes de la gravitation et de
  l'électromagnétisme relativité générale et théories unitaires}}.
\newblock Masson, 1955.

\bibitem{Tonnelat1955}
Marie-Antoinette Tonnelat.
\newblock {\em {La théorie du champ unifié d'Einstein et quelques-uns de ses
  développements}}.
\newblock Gauthier-Villars, 1955.

\bibitem{Tonnelat1966}
Marie~Antoinette Tonnelat.
\newblock {\em {Einstein’s Theory of Unified Fields}}.
\newblock Gordon and Breach, July 1966.

\bibitem{Synge1960}
John~L. Synge.
\newblock {\em {Relativity: The General Theory}}.
\newblock North-Holland Publishing, 1960.

\bibitem{Synge1956}
John~Lighton Synge.
\newblock {\em {Relativity: The Special Theory}}.
\newblock North-Holland Publishing Company, 1956.

\bibitem{Dirac1975}
Paul A.~M. Dirac.
\newblock {\em {General Theory of Relativity}}.
\newblock Wiley, 1975.

\bibitem{Fock1957}
Vladimir Fock.
\newblock {Three Lectures on Relativity Theory}.
\newblock {\em Reviews of Modern Physics}, 29(3):325--333, 1957.

\bibitem{Fock1955}
Vladimir~Aleksandrovich Fock.
\newblock {\em Теория пространства, времени и
  тяготения}.
\newblock State Publishing House for Technical-Theoretical Literature
  (Gostekhizdat), 1955.

\bibitem{Fock1959}
Vladimir~Aleksandrovich Fock.
\newblock {\em {The theory of space, time and gravitation}}.
\newblock Pergamon Press, 1959.

\bibitem{Martinez2020}
Jean-Philippe Martinez.
\newblock {\em {The Fock-Infeld Dispute: An Illustration of the Renaissance of
  General Relativity in the Soviet Union}}, pages 113--139.
\newblock Springer, 2020.

\bibitem{Fock1966}
Vladimir Fock.
\newblock {Grundprinzipien der Einsteinschen Gravitationstheorie}.
\newblock In Hans-Jürgen Treder, editor, {\em {Entstehung, Entwicklung und
  Perspektiven der Einsteinschen Gravitationstheorie}}, pages 26--38, Berlin,
  1966. Akademie-Verlag.

\bibitem{Sexl1962}
Roman~U. Sexl.
\newblock {Über die Interpretation der allgemeinen Relativitätstheorie von V.
  Fock}.
\newblock {\em Zeitschrift für Physik}, 169:343--352, 1962.

\bibitem{Rohrlich1963}
Fritz Rohrlich.
\newblock {The Principle of Equivalence}.
\newblock {\em Annals of Physics}, 22(2):169--191, May 1963.

\bibitem{Infeld1960}
Leopold Infeld and Jerzy Plebański.
\newblock {\em {Motion and Relativity}}.
\newblock Pergamon Press, 1960.

\bibitem{Kaiser2007}
David Kaiser.
\newblock {Turning physicists into quantum mechanics}.
\newblock {\em Physics World}, pages 28--33, May 2007.

\bibitem{Kaiser2002}
David Kaiser.
\newblock {Cold War requisitions, scientific manpower, and the production of
  American physicists after World War II}.
\newblock {\em Historical Studies in the Physical and Biological Sciences},
  33(1):131--159, 2002.

\bibitem{DeWitt1965}
Bryce DeWitt.
\newblock {\em {Dynamical Theory of Groups and Fields}}.
\newblock Gordon and Breach, 1965.

\bibitem{DeWitt1964}
Bryce DeWitt.
\newblock {Dynamical Theory of Groups and Fields}.
\newblock In Bryce DeWitt and Cécile DeWitt-Morette, editors, {\em
  {Relativity, Groups and Topology: Lectures Delivered at Les Houches During
  the 1963 Session of the Summer School of Theoretical Physics, University of
  Grenoble}}, pages 585--820. Gordon and Breach, 1964.

\bibitem{Anderson1967}
James~Leroy Anderson.
\newblock {\em {Principles of Relativity Physics}}.
\newblock Academic Press, 1967.

\bibitem{Witten1962}
Louis Witten, editor.
\newblock {\em {Gravitation: an Introduction to Current Research}}.
\newblock Wiley, 1962.

\bibitem{Mercier1979}
A.~Mercier, H.-J. Treder, and W.~Yourgrau.
\newblock {\em {On General Relativity: An Analysis of the Fundamentals of the
  Theory of General Relativity and Gravitation}}.
\newblock De Gruyter, December 1979.

\bibitem{Dicke1964}
Robert Dicke.
\newblock {\em {The Theoretical Significance of Experimental Relativity}}.
\newblock Documents on Modern Physics. Gordon and Breach, 1964.

\bibitem{ChoquetBruhat1968}
Yvonne Choquet-Bruhat.
\newblock {\em {Géométrie différentielle et systèmes extérieures}}.
\newblock Number~28 in Monographies Universitaires de Mathématiques. Dunod,
  Paris, 1968.

\bibitem{Robertson1968}
Howard~P. Robertson and Thomas~W. Noonan.
\newblock {\em {Relativity and Cosmology}}.
\newblock Saunders Physics Books. W. B. Saunders, Philadelphia, 1968.

\bibitem{Sciama1969}
Dennis~William Sciama.
\newblock {\em {The Physical Foundations of General Relativity}}.
\newblock Doubleday, 1969.

\bibitem{Rindler1969}
Wolfgang Rindler.
\newblock {\em {Essential Relativity: Special, General, and Cosmological}}.
\newblock Van Nostrand Reinhold, 1969.

\bibitem{Dicke1970}
Robert Dicke.
\newblock {\em {Gravitation and the Universe}}.
\newblock Number~78 in Memoirs. American Philosophical Society, 1970.

\bibitem{Peebles1971}
James Peebles.
\newblock {\em {Physical Cosmology}}.
\newblock Princeton University Press, 1971.

\bibitem{Zeldovich1971}
Yakov~Borisovič Zel'dovič and Igor~Dmitriyevič Novikov.
\newblock {\em {Relativistic Astrophysics Vol. 1: Stars and Relativity}}.
\newblock University of Chicago Press, 1971.

\bibitem{Weinberg1972}
Steven Weinberg.
\newblock {\em {Gravitation and Cosmology: Principles and Applications of the
  General Theory of Relativity}}.
\newblock John Wiley \& Sons, 1972.

\bibitem{Penrose1972}
Roger Penrose.
\newblock {\em {Techniques in Differential Topology in Relativity}}.
\newblock Society for Industrial and Applied Mathematics, January 1972.

\bibitem{Hawking1973}
Stephen~W. Hawking and George F.~R. Ellis.
\newblock {\em {The Large Scale Structure of Space-Time}}.
\newblock Cambridge University Press, 1973.

\bibitem{Treder1980}
Hans-Jürgen Treder, Horst-Heino von Borzeszkowski, Alwyn van~der Merwe, and
  Wolfgang Yourgrau.
\newblock {\em {Fundamental Principles of General Relativity Theories: Local
  and Global Aspects of Gravitation and Cosmology}}.
\newblock Springer, 1980.

\bibitem{Chandrasekhar1983}
S.~Chandrasekhar.
\newblock {\em {The Mathematical Theory of Black Holes}}.
\newblock Oxford University Press, 1983.

\bibitem{Straumann1981}
Norbert Straumann.
\newblock {\em {Allgemeine Relativitätstheorie und relativistische
  Astrophysik}}.
\newblock Springer, 1981.

\bibitem{Zeldovich1983}
Yakov~Borisovič Zel'dovič and Igor~Dmitriyevič Novikov.
\newblock {\em {Relativistic Astrophysics Vol. 2: The Structure and Evolution
  of the Universe}}.
\newblock University of Chicago Press, 1983.

\bibitem{Wald1984}
Robert Wald.
\newblock {\em {General Relativity}}.
\newblock University of Chicago Press, 1984.

\bibitem{Penrose1984}
Roger Penrose and Wolfgang Rindler.
\newblock {\em {Spinors and Space-Time Vol. 1: Two-Spinor Calculus and
  Relativistic Fields}}.
\newblock Cambridge Monographs on Mathematical Physics. Cambridge University
  Press, 1984.

\bibitem{Penrose1986}
Roger Penrose and Wolfgang Rindler.
\newblock {\em {Spinors and Space-Time Vol. 2: Spinor and Twistor Methods in
  Space-Time Geometry}}.
\newblock Cambridge University Press, 1986.

\bibitem{Schutz1985}
Bernard~F. Schutz.
\newblock {\em {A First Course in General Relativity}}.
\newblock Cambridge University Press, 1985.

\bibitem{Pirani1986}
Felix Pirani and Mike Crampin.
\newblock {\em {Applicable Differential Geometry}}.
\newblock Number~59 in London Mathematical Society Lecture Notes. Cambridge
  University Press, 1986.

\bibitem{Naber1989}
Gregory~L. Naber.
\newblock {\em {Spacetime and Singularities: An Introduction}}.
\newblock London Mathematical Society Student Texts. Cambridge University
  Press, 1989.

\bibitem{ChoquetBruhat1989}
Yvonne Choquet-Bruhat.
\newblock {\em {Graded Bundles and Supermanifolds}}.
\newblock Number~12 in Lecture Notes, Monographs and Textbooks in Physical
  Science. Bibliopolis, Neapel, 1989.

\bibitem{Blum_2018}
Alexander Blum, Roberto Lalli, and Jürgen Renn.
\newblock {Gravitational waves and the long relativity revolution}.
\newblock {\em Nature Astronomy}, 2(7):534--543, jun 2018.

\bibitem{Kuhn1962}
Thomas~S. Kuhn.
\newblock {\em {The Structure of Scientific Revolutions}}, volume~2 of {\em
  International Encyclopedia of Unified Science}.
\newblock University of Chicago Press, 1962.

\end{thebibliography}

%----------------------------------------------------------------------------------------

\end{document}